\documentclass[]{aa}
\usepackage{graphicx,epsf,psfig,epsfig,times,amsmath,amssymb}

\usepackage{natbib}
\bibpunct{(}{)}{;}{a}{}{,}

\def\xspec{{\sc xspec}}
\def\sax{{\it BeppoSAX}}

\def\lsim{\raisebox{-.5ex}{$\stackrel{<}{\sim}$}}
\def\gsim{\raisebox{-.5ex}{$\stackrel{>}{\sim}$}}

\begin{document}
\title{Spectral evolution of bright NS LMXBs with \textit{INTEGRAL}: an application of the thermal plus bulk Comptonization model}
%\title{Average hard X-ray emission from NS LMXBs: Observational evidence of different spectral states}

%   \subtitle{A systematic X-ray hard tail  - Radio connection}
   \author{L.I. Mainardi\inst{1}, A. Paizis\inst{1}, R.Farinelli\inst{2}, E. Kuulkers\inst{3},     J. Rodriguez\inst{4}, D. Hannikainen\inst{5}, P. Savolainen\inst{5},\\
   S. Piraino\inst{6,7}, A. Bazzano\inst{8}, A. Santangelo\inst{7}}
\offprints{A. Paizis, ada@iasf-milano.inaf.it}

\institute{
INAF-IASF, Sezione di Milano, Via Bassini 15, I--20133 Milano, Italy
\and Dipartimento di Fisica, Universit\`{a} di Ferrara, Via Saragat 1,I--44100 Ferrara, Italy 
\and ESAC, ISOC, Villa\~nueva de la Ca\~nada, Madrid, Spain
\and CNRS, FRE 2591, CE Saclay DSM/DAPNIA/SAp, F--91191 Gif sur Yvette Cedex, France
\and Mets\"ahovi Radio Observatory, TKK, Mets\"ahovintite 114, FI-02540 Kylm\"al\"a, Finland
\and INAF-IASF, Sezione di Palermo, Via Ugo La Malfa 153, 90146 Palermo, Italy 
\and IAAT, University of T\"ubingen, Sand 1, 72076 T\"ubingen, Germany
\and INAF-IASF, Sezione di Roma, Via del Fosso del Cavaliere 100, I-00133, Roma, Italy
}

\date{Received  / Accepted }

% \abstract{}{}{}{}{} 
% 5 {} token are mandatory
 
  \abstract
   {}
   {The aim of this work is to investigate in a physical and quantitative way the spectral evolution of bright 
   Neutron Star Low-Mass X-ray Binaries 
   (NS LMXBs), 
   with special regard to the transient hard X-ray tails.}
   {We analyzed \textit{INTEGRAL} data for five sources (GX~5--1, GX~349$+$2, GX~13$+$1, GX~3$+$1, GX~9$+$1) and built broad-band X--ray spectra from 
   JEM--X1 and IBIS/ISGRI data. For each source, X-ray spectra from different states were fitted %corresponding to a given hardness ratio range was fit 
   with the recently proposed model $\mathsf{compTB}$.}
   {The spectra have been fit 
     with a two-$\mathsf{compTB}$ model.  
   In all cases the first $\mathsf{compTB}$ describes the dominant part of the spectrum that we interpret as thermal Comptonization
  of soft seed photons ($<$1\,keV), likely from the accretion disk, by a 3--5\,keV corona. In all cases, this component does not evolve much in terms of 
    Comptonization efficiency, with the system converging to thermal equilibrium for increasing accretion rate. The second $\mathsf{compTB}$ varies 
    more dramatically spanning from
     bulk plus thermal Comptonization of blackbody
     seed photons to the blackbody
     emission alone. These seed photons (R$<$12\,km, kT$_{s}>$1\,keV), likely from the neutron star
     and the innermost part of the system, the Transition Layer, 
   are Comptonized by matter in a converging flow. The presence and nature of this second $\mathsf{compTB}$ component (be it a pure blackbody 
    or Comptonized) are related to the inner local accretion rate which can influence the transient behaviour 
    of the hard tail: high values of accretion rates correspond to an efficient Bulk Comptonization process (bulk parameter 
    $\delta\neq0$) while even higher values of accretion rates suppress the Comptonization, resulting 
   in simple blackbody emission ($\delta=0$).}
   {The spectral evolution of the sources has been successfully studied in terms of thermal and Bulk Comptonization efficiency, in relation to the physical conditions in the Transition Layer. %The reason for the transient nature of the hard tail as well as for its persistent absence in some sources are discussed.
    }
  
\keywords{Stars: individual: GX~5--1, GX~13$+$1, GX~3$+$1, GX~9$+$1, GX~349$+$2 -- X-rays: binaries -- binaries: close -- stars: neutron -- accretion}
\authorrunning{Mainardi et al.}
\titlerunning{Spectral evolution of bright NS LMXBs with \textit{INTEGRAL}}

\maketitle
%
%________________________________________________________________
%
\section{Introduction}
Low-Mass X-ray Binaries (LMXBs) are systems where a compact object, either 
a neutron star (NS) or a black hole candidate (BHC), accretes matter via Roche lobe
overflow from a normal companion of
mass \mbox{$\textit{M}\lesssim$ $1\textit{M}_{\odot}$}; 
a peculiar characteristic to this type of system is the
formation of an accretion disk in the orbital plane near the compact object.
In this paper we studied \textit{persistently bright} NS LMXBs ($L_X\approx 10^{37}-10^{38}$ erg sec$^{-1}$) granting long-lasting observability with instruments operating in the soft/hard X-ray range.
The spectra of these sources are usually described as the sum of two components \cite[e.g][]{mitsuda84,white88,barret01}: 
a soft component often likely associated with
the accretion disk or NS and a hard component 
interpreted as thermal Comptonization of soft seed photons from the disk and/or the NS by 
high temperature plasma of electrons near the compact object (so-called corona). 
The advent of broad-band X-ray missions, such as \textit{BeppoSAX, RXTE, INTEGRAL},
revealed the presence of a spectral \mbox{hardening} (so-called ``hard tails") above $\sim$30\,keV on top of otherwise soft spectra \cite[e.g][hereafter P06, and references therein]{disalvo02, paizis06}. 
These hard tails, mostly fit with phenomenological models such as a power-law, 
have been detected in Z sources \cite[e.g.][]{frontera98,disalvo02b} and also in 
the bright Atoll source GX~13$+$1 (P06). 

To explain the origin of these hard tails, different models have been
proposed  across the years, such as direct synchrotron emission from a jet \citep{markoff05}, 
hybrid thermal/non-thermal Comptonisation \citep{coppi99,disalvo06} or, more recently, bulk motion Comptonisation \citep[][hereafter TMK97 and F08, respectively]{titarchuk97,farinelli08}.
The adoption of different models which satisfactorily  fit the hard X--ray 
component in NS LMXBs clearly shows that it is necessary to look for 
other observable quantities in these systems, such as their timing
properties or correlation among the spectral parameters.

For black-hole sources, a fundamental step forward came with the unambiguous
discovery of the saturation of the spectral index (as a function of either
the low QPO frequency or seed photon BB normalization) when  sources moves
from the low/hard to the high/soft state \citep[e.g.][and references therein]{shaposhnikov09}.
Such a saturation can be naturally explained in terms of the
presence 
of a converging flow close to the compact object.

For NS systems, the situation is more complex because of the presence of a firm surface in the compact object that plays a fundamental role in determining the hydrodynamical and
radiative system configuration with, in turn, a different spectral appearance.
However, given that a PL-like emission can be seen in both BH and NS systems
and on the basis of the above reported results for BHs, it is a reasonable 
working hypothesis that converging flow (with unavoidable qualitative
and quantitative differences) is a characteristic of NS systems too, at
least in some particular accretion rate states.
Close to the NS surface, the radial component of the velocity of matter ($V_{R}$) dominates 
over the azimuthal one and it is straightforward to 
find that $V_{R}$ $\sim$0.56 c at $R_{NS}$, factor that is actually reduced by 
the presence of pressure gradients (due to radiation, gas and magnetic 
field).
Our interpretation is that multiple thermal plus bulk Compton 
scatterings, that occur in a high optical depth environment, are at the 
origin of the hard tails observed in NS LMXBs, extending over the thermal 
continuum up to at least 100\,keV.
A self-consistent physical treatment of the innermost region in the case of neutron stars is presently being developed by  Titarchuk \& Farinelli (in preparation, see Sect.~\ref{transient}). In this paper we adopt
the bulk motion Comptonization scenario which also enables a homogeneous interpretation 
with previous works \citep[e.g. P06,][]{farinelli07,farinelli08,farinelli09}.

\begin{figure}
\centering
\includegraphics[width=1.0\linewidth]{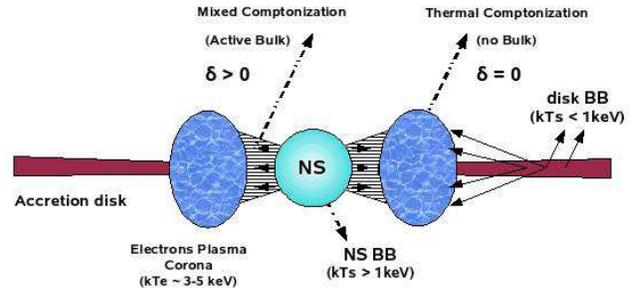}
%\vspace{4cm}./image_ticks.ps
\caption{Schematic view of the proposed scenario for thermal and bulk Comptonization 
regions in LMXBs hosting NSs.
\label{fig:TL}}
\end{figure}

P06 proposed for the first time a 
unified physical scenario to explain the spectral evolution of NS LMXBs, including the peculiar transient hard tail.
All the observed spectral states could be well fit in terms of the interplay of 
thermal and bulk Comptonization (TC and BC, respectively) using the $\mathsf{BMC}$ model \cite[][hereafter TMK96 and TMK97]{titarchuk96,titarchuk97} in \xspec. 
Thermal Comptonization operates in a region where a hot plasma is present and acts on a population 
of soft seed photons, hardening the injected seed photon spectrum (from the accretion disk, Fig.~\ref{fig:TL}); 
the resulting spectrum is extended to higher energies depending on the Comptonization efficiency. Bulk Comptonization operates in the inner part of the system between the Keplerian accretion disk and the NS surface, the Transition Layer, hereafter TL (Fig.~\ref{fig:TL}). 
In the inner part of the TL, matter falls towards the NS surface 
with a preferred (bulk) motion, acting on the inner seed photons of the system; the result is a power-law-shape spectrum at high energies due to the energy transfer from the matter kinetic 
energy to the photon field (TMK96, TMK97). P06 found evidence that bulk motion is always present and its strength is 
related to the accretion rate and is suppressed only in the presence of high local luminosity. This scenario has been confirmed also by \cite[][hereafter F07]{farinelli07},
who re-analyzed broad-band \sax\ data, applying the $\mathsf{BMC}$ model to \mbox{GX~17$+$2}. 

In the quest to study in a \emph{quantitative} way the evolution of the parameters describing the innermost physical conditions of NS LMXBs, \cite{farinelli08} (hereafter F08), developed a new Comptonization model, 
$\mathsf{compTB}$\footnote{http://heasarc.gsfc.nasa.gov/docs/xanadu/xspec/models/comptb.html}. 
We recall here the basics of this recent model, referring the reader to F08 for a more detailed description. 

The total emerging spectrum is given by
\begin{equation}
%F(E)=\frac{C_n}{A+1}(BB + A \times BB \ast G)\label{compTB}
F(E)=\frac{C_n}{A+1}(\underbrace{BB}_{(a)} + \underbrace{A \times BB \ast G}_{(b)})\label{compTB}
\end{equation}
where $C_n$ is the normalization constant, $A$ is the illumination factor, $C_n\,BB/(A+1)$ 
is the seed photon spectrum directly seen by the observer and not modified by Comptonization processes.
 $C_n\,A/(A+1)\times BB \ast G$ is the Comptonized spectrum obtained by convolution of a seed photon spectrum 
(blackbody, BB) with the Green's Function\footnote{The Green's Function, $G$, is the response of the system to the injection of a 
monochromatic line, see F08 for details.}, in order to evaluate the effect of TC and BC on the seed photon field.

Parameters of the $\mathsf{compTB}$ \xspec\ model are the seed photon temperature $kT_s$, 
the electron plasma temperature $kT_e$, the spectrum energy slope (i.e. overall Comptonization efficiency) 
$\alpha$ (photon index $\Gamma=\alpha+1$), the bulk parameter 
$\delta$ that quantifies the efficiency of BC  over  TC and $\log A$, which assigns a different 
weight to the two components $(a)$ and $(b)$. This model enables the co-existence of the direct seed photon
component and its Comptonized part, all obtained in a self-consistent way. In Eq.[\ref{compTB}] we note that 
for $\log A = -8$, we have only $(a)$, i.e. the direct seed photon component, while for $\log A = 8$, 
the direct component is no longer
visible and we have only thermal plus bulk Comptonization. In the case of  
$\delta=0$, the BC contribution is neglected, retaining
%%maintaining 
only the TC effects \cite[equivalent to compTT,][]{titarchuk94}.

The first application of the $\mathsf{compTB}$  model and of the proposed physical scenario to the spectral evolution of a single NS LMXB, to trace in a quantitative way the evolution of the physical parameters, was
presented in \cite{farinelli09} (hereafter F09) on Cyg~X--2, using \sax\ data. F09 fitted the spectra of the source with two $\mathsf{compTB}$ models. 
The first one, with bulk parameter $\delta$=0, was the dominant component and described a pure thermal 
Comptonization process. This component, due to the Comptonization 
of cold disk photons by the \emph{outer} TL, is roughly constant with time 
 and the accretion rate. 
The second $\mathsf{compTB}$  model
describes the overall Comptonization (thermal plus bulk, variable $\delta$ parameter) 
of hotter seed photons close to the neutron star surface by the \emph{inner} region 
of the TL (Fig.~\ref{fig:TL}). This component is highly variable with time (and local accretion rate)
ranging from significant
Comptonization of the hot seed photons (thermal and bulk) to a simple blackbody-like 
spectrum.

In this work, we extend the above study to five persistently bright NS LMXBs (GX~5--1, GX~349$+$2, 
GX~13$+$1, GX~3$+$1 and GX~9$+$1). 
The choice of sources was driven 
by the validity range of the $\mathsf{compTB}$ model (diffusion regime, see
F08), associated to a high accretion rate where the transient hard tail can be 
appreciated. The selected sources are located in the Galactic Centre
where, because of the presence of a large number of  X--ray sources, 
angular resolution is an issue. The  imager IBIS \citep{ubertini03} and 
the X-ray instrument JEM--X \citep{lund03} on-board \textit{INTEGRAL} \citep{winkler03}
have the necessary angular resolution to disentangle the different contributions, 
allowing to build good quality broad-band X-ray spectra (3--200\,keV).
\begin{figure*}
\centering
\includegraphics[width=1.0\textwidth]{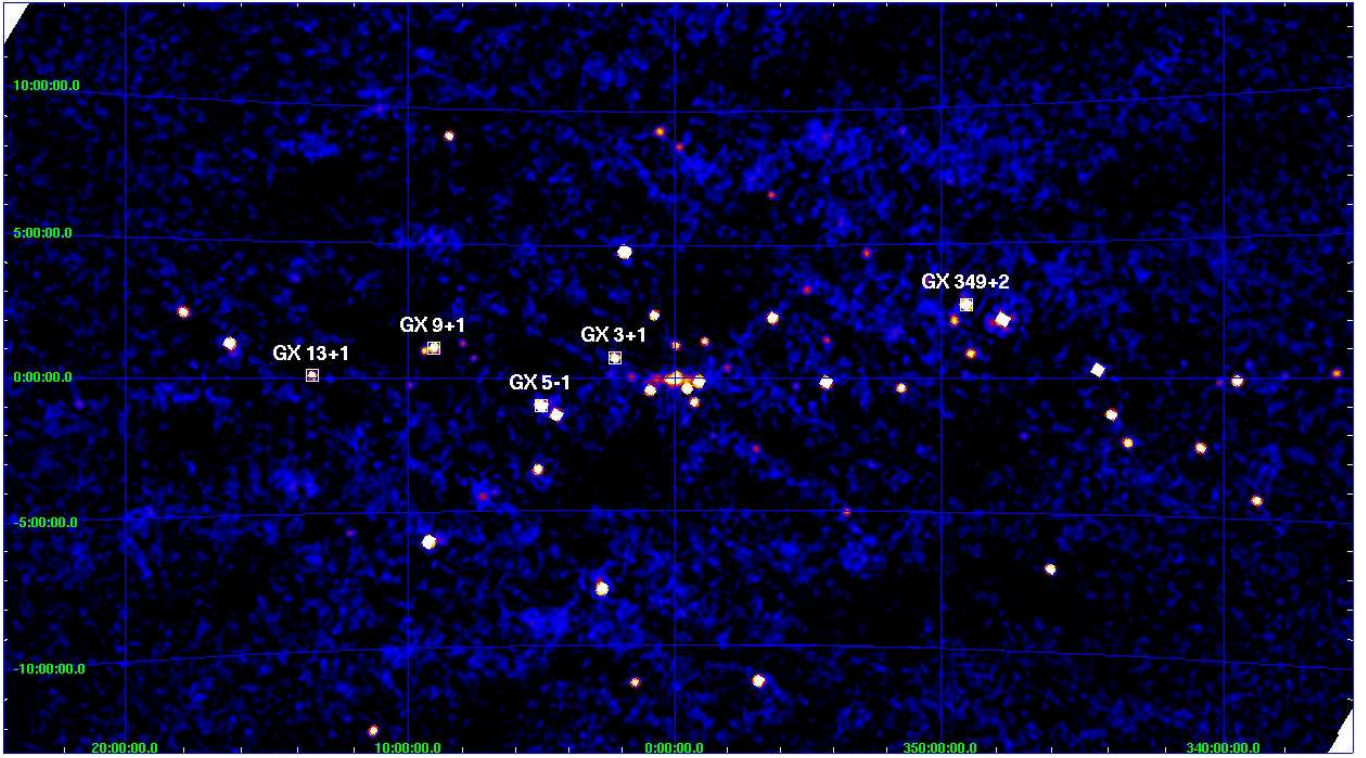}
%\includegraphics[width=1.0\linewidth]{/nereide/laura/gc_isgri_20-30keV_1.ps}
%%\vspace{4cm}./image_ticks.ps
\caption{IBIS/ISGRI 20--40\,keV mosaic in Galactic coordinates of the Galactic Bulge (about 3.3\,Msec).
Only the sources studied in this paper are labeled. 
\label{fig:mosaic}}
\end{figure*}

%\citep{winkler03}.
%
\begin{table}
  \begin{center}

    \caption{The five Bright NS LMXBs studied in this paper. References for distance 
    and $N_{H}$: \cite{grimm02} and \cite{ueda05} (GX~5--1), \cite{grimm02} and \cite{iaria04b} (GX~349$+$2), \cite{bandyopadhyay99} and \cite{ueda04} (GX~13$+$1), \cite{kuulkers00} and \cite{chevenez06} (GX~3$+$1), \cite{iaria05} (GX~9$+$1).}\vspace{1em}
    \renewcommand{\arraystretch}{1.2}
    \begin{tabular}[h]{cccc}
      \hline
Source &  D$^a$ & $N_{H}^b $ & pointings \#  \\
\hline
\hline
Z sources &	\\
\hline
GX~5--1   & 7.2& 2.8 &  137   \\
GX~349$+$2   &9.2 & 0.77 &  51    \\ 
\hline
Atoll sources &\\
\hline
GX~13$+$1 & 7& 3.2 &  35   \\
GX~3$+$1& 5 &1.66  &  166   \\
GX~9$+$1& 5 & 0.8 &  87 \\
\hline
\multicolumn{4}{l}{$^a$\quad The distance is in kpc.}\\
%\multicolumn{4}{l}{GX~5--1: Grimm et al. 2002.}\\
%\multicolumn{4}{l}{GX~349$+$2: Grimm et al. 2002.}\\
%\multicolumn{4}{l}{GX~13$+$1: Bandyopadhyay et al. 1999.}\\
%\multicolumn{4}{l}{GX~3$+$1: den Hartog et al. 2003.}\\
%\multicolumn{4}{l}{GX~9$+$1: Iaria et al. 2005.}\\
\multicolumn{4}{l}{$^b$\quad In units of $10^{22}$ cm$^{-2}$.}\\
%\multicolumn{4}{l}{GX~5--1: Ueda et al. 2005.}\\
%\multicolumn{4}{l}{GX~349$+$2: Iaria et al. 2004.}\\
%\multicolumn{4}{l}{GX~13$+$1: Ueda et al. 2004.}\\
%\multicolumn{4}{l}{GX~3$+$1: Chevenez et al. 2006.}\\
%\multicolumn{4}{l}{GX~9$+$1: Iaria et al. 2005.}\\
      \end{tabular}      
    \label{tab:table1}
  \end{center}
\end{table}
\section{Observations and data analysis}
 \subsection{Analysis process}
 The sample of the NS LMXB chosen is given in Tab.~\ref{tab:table1} and comprises five bright (``GX'') bulge sources. We have analyzed \textit{INTEGRAL}  pointings collected from September 2006 to September 2007 
 from
 the  AO4 and AO5 \textit{INTEGRAL} Key Programmes. The exposure time for each pointing is variable ($\Delta t \sim 2000\,$sec). 
 The selected pointings satisfy the condition that every source is $\leq 3^{\circ}$ from the pointing direction.
 Analysis was
performed on JEM-X1 and IBIS/ISGRI \citep{lebrun03} data with the software package OSA, version 7.0.

 The energy bands selected for JEM-X1 imaging analysis are: 3--10\,keV and 10--20\,keV  
 while the energy band selected for IBIS/ISGRI imaging analysis is: 20--40\,keV. The mosaic of the Galactic Centre (Fig.~\ref{fig:mosaic}) 
 was constructed using all 
 available pointings in the IBIS/ISGRI energy band: 
 as expected, all the sources studied were active and detectable.
 We 
 used the official \textit{INTEGRAL} catalog to extract spectra for the single sources and the JEM-X default response matrix (256 bins) 
 and a new re-binned (57-bins) IBIS/ISGRI response matrix.
  \subsection{Spectral classification with HR}
  In order to study the spectral evolution of the sources, 
  we associated a numeric 
  parameter to the spectrum of each pointing, the \textit{Hardness Ratio} (hereafter HR). 
  The HR is computed from the counts per second extracted from the two energy bands of JEM-X analysis
  as follows: 
  \begin{equation}
HR= \frac{S}{H}\label{HR}
\end{equation}
where S\,(soft) is the count rate for the 3--10\,keV energy band while H\,(hard) is the count rate for the 
10--20\,keV energy band. HR is an indicator of the source spectral state: the lower the HR the harder the spectrum. 
Despite its unusual/counterintuitive form, we adopted the definition~(\ref{HR}) because the good statistics showed by JEM-X granted us integer values for a more hand-light management of HR boundaries and, as explained in section~\ref{transient}, we wanted to connect the increase of HR with the increase of the accretion rate.

\begin{figure}
\centering
\includegraphics[angle=90,width=1.0\linewidth]{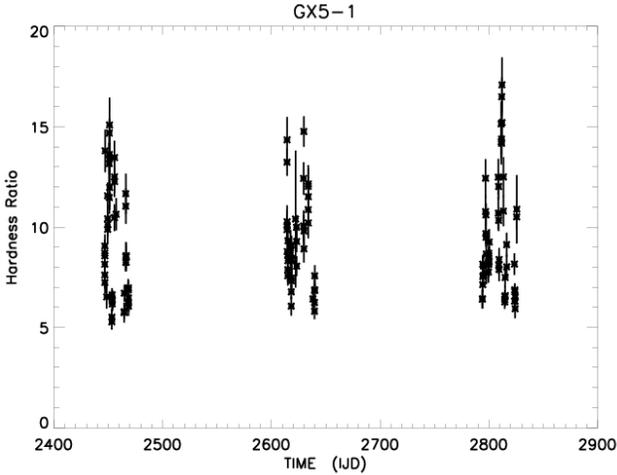}
\caption{Example of the HR variation for the source GX 5--1 considering all available pointings.
\label{fig:GX5-1_HR}}
\end{figure}

Considering the overall behaviour, each source presents a characteristic HR variability 
spanning wide ranges (for example see Fig.~\ref{fig:GX5-1_HR} for GX~5--1). From a morphological evaluation of the spectral characteristics together with HR values, 
we classified every spectrum and averaged similar spectra to improve the overall statistics. %giving us a different number of HR intervals for each source. 
%Details are given in the sections related to individual sources.
%  
 \subsection{The spectral fitting}
 Before spectral fitting, all the spectra were
 %%have been 
 re-binned so as to contain at least 25 counts in every channel; we considered only the counts within the interval 5--18\,keV for JEM-X1 spectra and the counts starting
 at 22\,keV for IBIS/ISGRI. 
Systematic errors of $2\%$ for IBIS/ISGRI data and $3\%$ for JEM-X1 data were added.
Instrumental cross-calibration factors 
%%have been 
were allowed to vary between 0.8 and 1.2 (IBIS/ISGRI) with JEM-X1 frozen at 1. In all cases the value of the constant stabilized itself at the maximum value,  1.2. For the spectral analysis, we used \xspec\ version 12.

% We have used the \xspec\ program, version 12, to accomplish the spectral analysis.\\ 
 In all cases we started the fitting process with the simplest general model considering only one $\mathsf{compTB}$  and then in more complex cases, we added a second $\mathsf{compTB}$, as in F08.
 Hence (excluding the cross-calibration constant), in the most complex cases, we used:
 \begin{equation}
F(E)=\mathsf{wabs}\ast(\mathsf{compTB[1]} + \mathsf{compTB[2]}).
\end{equation}
where  $\mathsf{wabs}$ \citep{morrison83} is the multiplicative model that takes into account the interstellar absorption at low X-ray energies; 
the column-density, $N_H$, was
%%has been 
 fixed during the fitting 
 to the values shown in Table~\ref{tab:table1}, as the absence of data below 5\,keV 
 does not allow the
correct determination of this parameter through the fitting process.

We note that $\mathsf{compTB[1]}$, with parameters set to  $logA=8$ and $\delta=0$, is a \textit{thermal} $\mathsf{compTB}$ accounting only for TC processes 
(equivalent to $\mathsf{compTT}$). Instead, $\mathsf{compTB[2]}$, when present,
is either
 a \textit{mixed} $\mathsf{compTB}$ (shape-wise equivalent to $\approx \mathsf{bbody}+ \mathsf{PL}$, where we consider both TC and BC processes), or  
 a $\mathsf{bbody}$ (neither TC nor BC setting $\log A=-8$).
For  the cases in which the fit required 
two $\mathsf{compTB}$ components, a single temperature of the plasma $kT_e$ was considered (see section~\ref{results} for details).  
%because a preliminary fit showed a convergence for this parameter. %the data do not require two different ones .  
%
 \subsection{Other parameters}\label{other}
Some other quantities can be obtained from the fit processes in order to have more information about the system and the environment around the NS:  
\begin{itemize}
\item $\mathbf{F_{tot}}$: the unabsorbed total flux (erg sec$^{-1}$ cm$^{-2}$) in the interval 0.1--200\,keV and relative 
luminosity $L_{tot}=4\pi D^2 F_{tot} $ (using distances in Table~\ref{tab:table1});
\item CAF: the \textit{Compton Amplification Factor} that allows us to evaluate the total energy gain of the seed 
photons through the TC process. It is defined as $CAF=flux1/flux2$, where \textit{flux1} (erg sec$^{-1}$ cm$^{-2}$) is the unabsorbed energy flux associated to the
thermally Comptonized photons (calculated on \textsf{compTB[1]} with $N_H=0$), and \textit{flux2} (erg sec$^{-1}$ cm$^{-2}$) is the unabsorbed energy flux associated to
the seed photons (calculated on \textsf{compTB[1]} with $\log A=-8$ and $N_H=0$) in the energy interval 0.1--200\,keV;
\item The optical depth $\mathbf{\tau}$: the Comptonization parameter $\mathbf{\alpha}$ from the fit is linked to $kT_e$ and $\tau$ as: %shown in Titarchuk \& Lyubarskij (1995) that, in \textit{no relativistic regimes}, reduces to:
\begin{equation}
\alpha=-\frac{3}{2}+\sqrt{\frac{9}{4}+\frac{\pi^2m_ec^2}{C_{\tau}kT_e (\tau + \frac{2}{3})^2}}\label{alpha}
\end{equation} 
where $C_{\tau} =3$ for a spherical geometry with optical radius $\tau$, while $C_{\tau} =12$ for a ``slab'' geometry \citep{titarchuk95}. In this paper we presumed 
a \textit{slab geometry} corresponding to a not very extended corona (see also F08); 
\item $\mathbf{R_{bb}}$: the blackbody radius of $\mathsf{compTB[2]}$, computed using $L_{bb}=4\pi\,R_{bb}^2\,\sigma T^4$ .
\end{itemize}
 \section{Results}\label{results}
 The detailed results per source are given in the next subsections while here we note a few general trends and issues. Each source shows
  %%has shown 
  a given range of variability for HR values which enabled
  %%brought 
  us to identify 
 a different number of intervals (see Table~\ref{tab:tableGX_all}) and spectral states per source .
 Our sources showed all types of behaviours, and hence  
%%Spanning over all the encountered cases, 
we used a model with: two $\mathsf{compTB}$s in which the first is thermal and the second  
\textit{mixed}  (GX~5--1 and GX~13$+$1); two $\mathsf{compTB}$s in which the first is thermal and the second a \textit{blackbody} (GX~9$+$1 and GX~349$+$2);
 one thermal $\mathsf{compTB}$ (GX~3$+$1).
The absence of data below 5\,keV  %%has 
led to a high uncertainty in determining $kT_s$ 
%%determination 
in the $\mathsf{compTB[1]}$ component (the softest). 
In fact, for the majority of cases only an upper limit could be determined.

As already experienced in F08 and F09, in the cases where we applied a \textit{mixed} $\mathsf{compTB[2]}$, the lack of an observable 
cut-off prevented us to allow  
$\alpha$, $\delta$ and $\log A $ to be free during the fit: indeed if the number of the free parameters exceeds the number of the observable quantities (normalization, cut-off and the PL slope), a degeneration of model parameters 
occurs and hence more than one model solution can be obtained. The degeneration 
can be reduced: as in F08, the $kT_e$ value for $\mathsf{compTB[2]}$ was set equal to the $\mathsf{compTB[1]}$ one, considering a single plasma temperature (indeed a preliminary fit provided similar values for both $kT_e$). Furthermore, we fixed $\log A$ (\mbox{GX~5--1)} or $\alpha$ (\mbox{GX~13$+$1}) to the values corresponding to a $\chi^2$ minimum (chosen inside an interval calculated with the \xspec\ command \emph{steppar}). We left $\delta$ free during the fit to evaluate the bulk efficiency.\\
\begin{table}
  \begin{center}

    \caption{The number of pointings and exposition time for each HR interval for the NS LMXBs studied in this paper.          }\vspace{1em}
    \renewcommand{\arraystretch}{1.2}
    \begin{tabular}[h]{cccc} 
      \hline
Interval & pointings  &\multicolumn{2}{c}{T$_{exp}$ (ksec)}\\
	& N& JEM-X & ISGRI	\\	
\hline
\hline
\multicolumn{4}{c}{GX~5--1}\\
\hline
HR$<$7   &36 & 102 & 69     \\ 
7$<$HR$<$10  &52 & 155 &  106   \\
10$<$HR$<$12 &23  & 58 & 38    \\
HR$>$12 &26 & 80 & 53   \\
\hline
\multicolumn{4}{c}{GX~13$+$1}\\
\hline
HR$<$10   &11 &29  &18      \\ 
10$<$HR$<$13  &10 & 26 & 17    \\
HR$>$13 &14  & 35 & 23   \\
\hline
\multicolumn{4}{c}{GX~9$+$1}\\
\hline
HR$<$7.5   & 19 & 50  & 34     \\ 
7.5$<$HR$<$8.5  & 47 &119  & 79    \\
HR$>$8.5 & 21  & 54  & 37   \\
\hline
\multicolumn{4}{c}{GX~349$+$2}\\
\hline
HR$<$7   & 20& 52 &  35    \\ 
HR$>$7 & 31 & 82 &  53  \\
\hline
\multicolumn{4}{c}{GX~3$+$1}\\
\hline
HR$<$8   &37 & 107 & 70     \\ 
8$<$HR$<$9.5  &103 & 323 & 210    \\
HR$>$9.5 & 26 & 78 & 52   \\
\hline

      \end{tabular}
    \label{tab:tableGX_all}
  \end{center}
\end{table}
%------------------------------
\subsection{GX~5--1}
The variability range of HR (definition [\ref{HR}]) spans between $\sim$5--18 (as shown in Fig.~\ref{fig:GX5-1_HR}). 
By classifying the 
spectral shapes, we obtained four mean spectra %(HR$<$7, 7$<$HR$<$10, 10$<$HR$<$1, HR$>$12) %(Table~\ref{tab:tableGX5-1})
: the first spectrum, the hardest one, corresponding to  HR$<$7, shows, as expected, a hard tail between 40 and 150\,keV. Indeed a hard tail in the spectra of GX~5--1
had already been detected by previous observations \citep[][ P06]{asai94, paizis05}. The second spectrum (7$<$HR$<$10) shows a weaker hard tail between 40 and 150\,keV; the third one (10$<$HR$<$12) does not show a hard tail and is soft; the last spectrum is the softest (HR$>$12), again with no hard tail \citep[see also][on the variability of GX~5--1]{kuulkers07, paizis05}.

\begin{figure*}

\hbox{\hspace{0.5cm}
\includegraphics[scale=0.35]{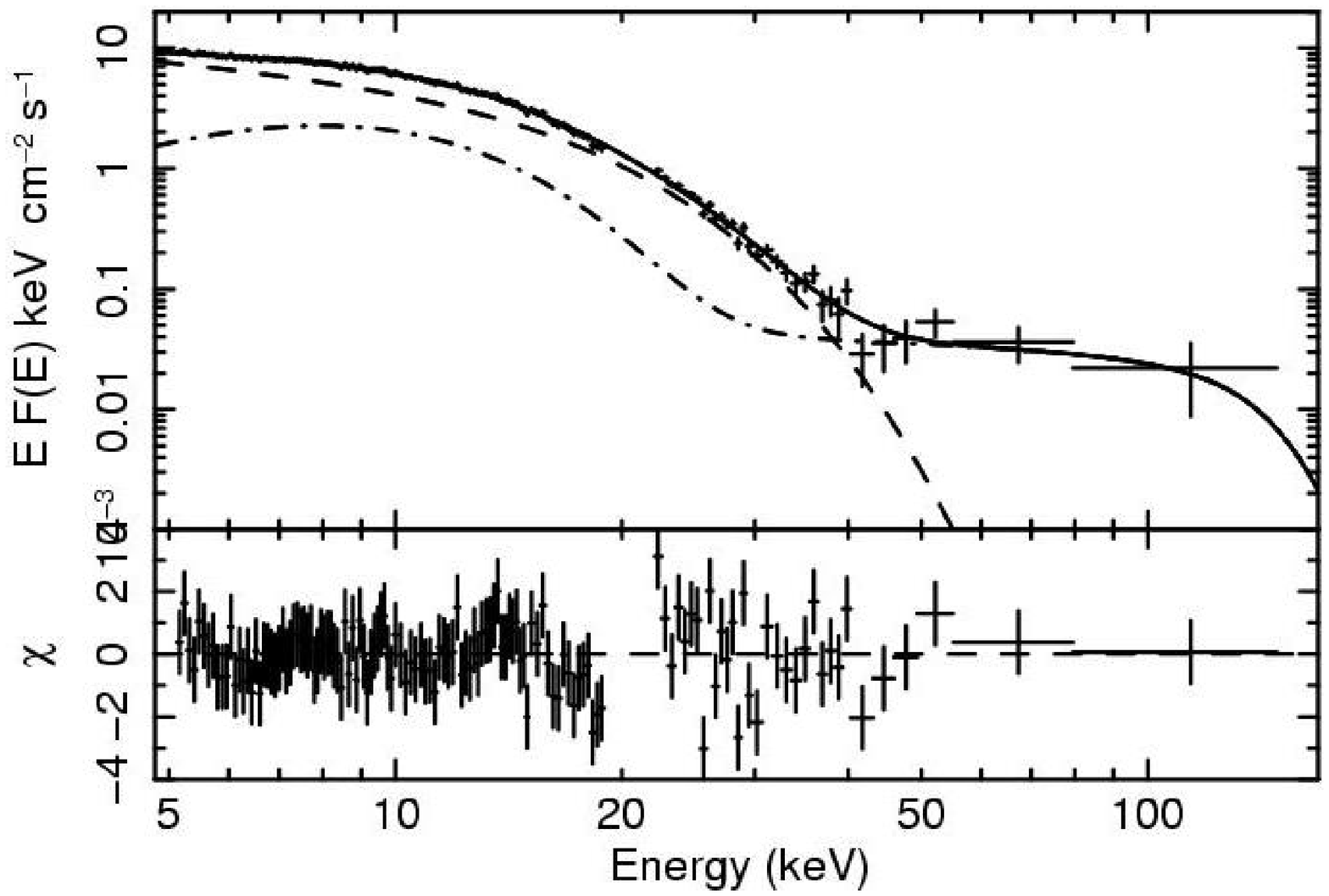}
\hspace{0.5cm}
\includegraphics[scale=0.35]{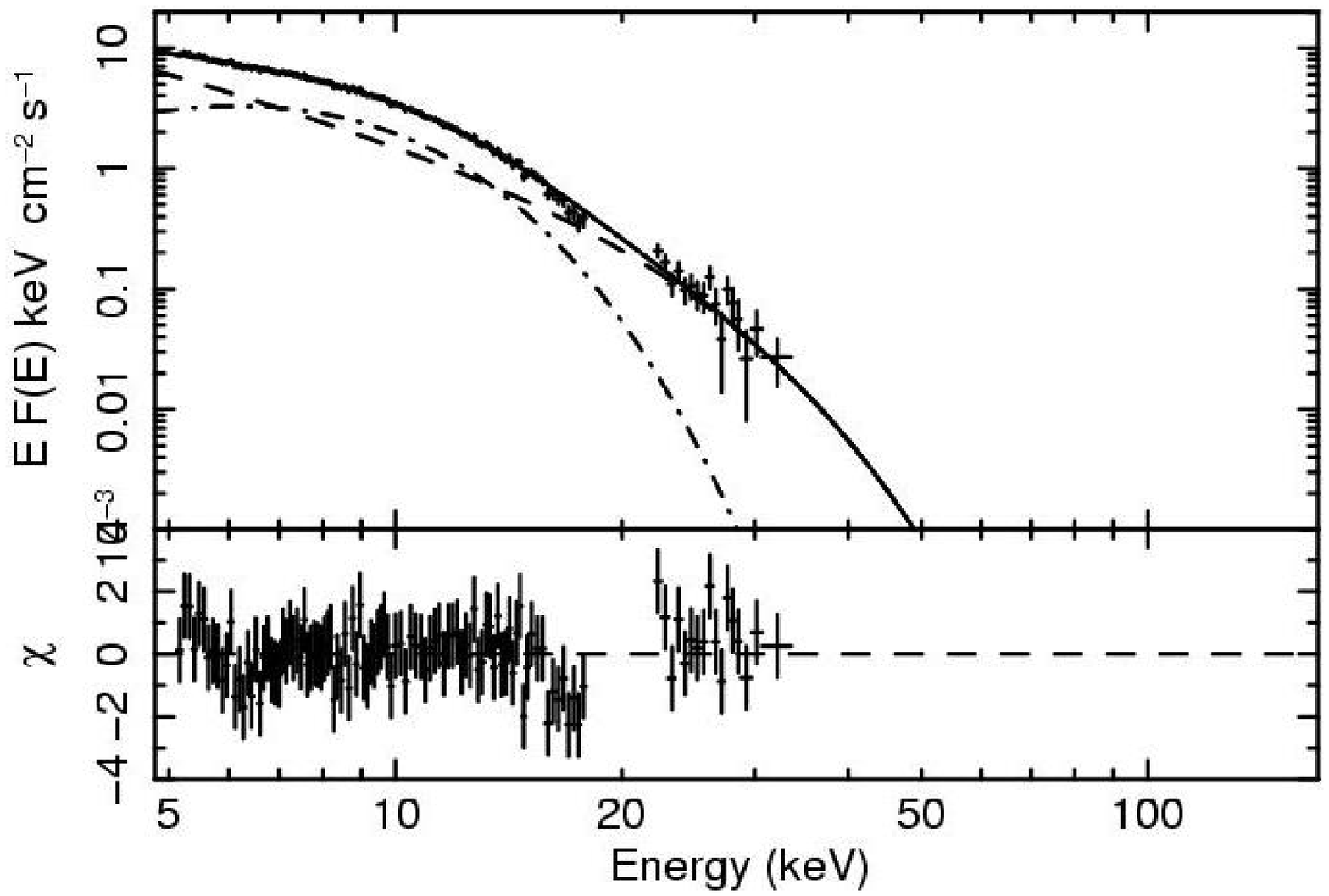}}
%\vbox{\vspace{0.15cm}}
%\hbox{\hspace{0.5cm}
%\includegraphics[scale=0.35]{./gx5_3_eeuf.ps}
%\hspace{0.5cm}
%\includegraphics[scale=0.35]{./gx5_4_eeuf.ps}}
%\vbox{\vspace{0.15cm}}
%\hbox{\hspace{0.5cm}
%\includegraphics[height=7.5cm,angle=-90]{}
%\hspace{0.5cm}
%\includegraphics[height=7.5cm,angle=-90]{}}
\caption[]{Unabsorbed EF(E) spectra of GX~5--1 for the HR intervals: HR$<$7 (left), and HR$>$12 (right). The dotted lines are the two separated $\mathsf{compTB}$  components: dash for  $\mathsf{compTB[1]}$, dash-dot  $\mathsf{compTB[2]}$.}
%One Crab is approximately 93\,counts/sec in IBIS/ISGRI 
%22--40\,keV and 61\,counts/sec in IBIS/ISGRI 40--80\,keV. 
\label{fig:GX5-1eeuf}
\end{figure*}

As required by the data, we used the two-component model for all spectra in the fitting process and obtained the following results (see Table~\ref{tab:tableGX5-1_par}):
\begin{itemize}
\item with the $\mathsf{compTB[1]}$ component, we describe the presence of the TC process as the dominant mechanism in all the four spectra. For higher values of HR, we observe that the CAF decreases, while $\alpha$ increases. This is consistent with a reduction in the efficiency of the TC process;
\item the seed photon temperatures $kT_s$ of the two components are different: in 
the \textit{thermal} $\mathsf{compTB[1]}$, $kT_s$ remains always below 1\,keV while in the \textit{mixed} $\mathsf{compTB[2]}$ it has 
higher values $\gtrsim 1.5$\,keV; %this difference could be associated to two separated emission regions at different temperatures; 
\item considering the errors associated, the plasma temperature $kT_e$ remains within the interval 3--4\,keV;
\item with $\mathsf{compTB[2]}$ in all spectra, we describe an additional component that shows a specific evolution: in the first two spectra (HR$<$7, 7$<$HR$<$10) we obtained 
a \textit{mixed} $\mathsf{compTB}$  that  allows us to evaluate the BC contribution with  the parameter $\delta\neq0$. 
The spectra are representative of the 
%%We are in the presence of an 
\textit{intermediate} state (according to the definition given by P06). For increasing values of HR, $\log A$ 
%%has shown 
shows a decreasing trend that assigns less importance to the mixed component. In fact, in the third and fourth spectra (10$<$HR$<$12, HR$>$12), we found a  best-fit using the  $\mathsf{compTB[2]}$ in the blackbody mode ($\log A=-8$ ); 
\item with increasing values of HR, the size of the 
blackbody-emitting region,
$R_{bb}$ in $\mathsf{compTB[2]}$, increases, remaining compatible with the NS and/or TL dimensions.
\end{itemize}
In Fig.~\ref{fig:GX5-1eeuf} we show 
the EF(E) spectra (keV\,cm$^{-2}$\,sec$^{-1}$) of GX~5--1 for two of the four selected HR intervals (corresponding to the hardest and the softest case). 
%
%\begin{table}
%  \begin{center}
%
%    \caption{The bright NS LMXB GX~5--1          }\vspace{1em}
%    \renewcommand{\arraystretch}{1.2}
%    \begin{tabular}[h]{cccc} 
%      \hline
%Interval & pointings  &\multicolumn{2}{c}{T$_{exp}$ (ksec)}\\
%	& N& JEM-X & ISGRI	\\	
%\hline
%\hline
%HR$<$7   &36 & 102 & 69     \\ 
%7$<$HR$<$10  &52 & 155 &  106   \\
%10$<$HR$<$12 &23  & 58 & 38    \\
%HR$>$12 &26 & 80 & 53   \\
%\hline
%
%      \end{tabular}
%    \label{tab:tableGX5-1}
%  \end{center}
%\end{table}
%
\begin{table}
\begin{center}
\caption{Best-fit parameters of the multi-component model $\mathsf{wabs\cdot(compTB[1]+compTB[2])}$ for the source GX~5--1. Errors are computed at 90$\%$ confidence level for a single parameter.}
\begin{tabular}[h]{ccccc}
\hline
\hline
parameter&HR$<$7&7$<$HR$<$10&10$<$HR$<$12&HR$>$12\\
\hline
\multicolumn{5}{c}{$\mathsf{compTB[1]}$ (\,thermal: $\log A=8$, $\delta=0$\,)}\\
\hline
$kT_s$ $^a$&$0.4\,(<0.6)$&$0.6\,(<0.7)$&$0.4\,(<0.6)$&$0.4^{+0.1}_{-0.2}$\\
$kT_e$ $^a$&$3.45^{+0.05}_{-0.14}$&$3.11^{+0.05}_{-0.13}$&$3.7^{+0.3}_{-0.5}$&$[4]$\\
$\alpha$&$1.5^{+0.1}_{-0.1}$&$1.5^{+0.1}_{-0.2}$&$2.1^{+0.2}_{-0.5}$&$2.5^{+0.1}_{-0.1}$\\
$\tau$ $^d$&3.6&3.8&2.6&2.1\\
CAF $^d$&1.6&1.5&1.4&1.3\\
\hline
\multicolumn{5}{c}{$\mathsf{compTB[2]}$}\\
\hline
$kT_s$ $^a$&$2.02^{+0.03}_{-0.11}$&$1.6^{+0.1}_{-0.1}$&$1.71^{+0.04}_{-0.04}$&$1.57^{+0.02}_{-0.02}$\\
$R_{bb}$ $^{b,d}$&$3.8^{+0.5}_{-0.1}$&$6.2^{+1.2}_{-0.8}$&$5.6^{+0.3}_{-0.2}$&$7.6^{+0.2}_{-0.2}$\\
$kT_e$ $^a$&$[kT_e]$&$[kT_e]$&-&-\\
$\alpha$&$1.2^{+0.9}_{-0.2}$&$1.1^{+0.4}_{-0.3}$&-&-\\
$\delta$&$36\,(>18)$&$33\,(>14)$&-&-\\
$\log A$&$[-1.82]$&$[-2.19]$&$[-8]$&$[-8]$\\
\hline
%$L_{tot}$ $^{c,d}$&2.2&2.3&3.0&2.9\\ 
%\hline
%$L_{bb}$ $^{c,d}$&0.3&0.3&0.4&0.5\\
$L_{tot}$ $^{c,d}$&2.5&2.5&3.5&3.5\\ 
\hline
$L_{bb}$ $^{c,d}$&0.3&0.3&0.4&0.5\\
&(12$\%$)&(12$\%$)&(11$\%$)&(14$\%$)\\
\hline
$\chi^2/dof$&133/130&101/118&82/114&96/109\\
\hline
\\
\multicolumn{5}{l}{$^a$\quad In keV.}\\
\multicolumn{5}{l}{$^b$\quad In km.}\\
\multicolumn{5}{l}{$^c$\quad In units of $10^{38}$\,erg\,sec$^{-1}$, in the 0.1--200\,keV energy range.}\\
\multicolumn{5}{l}{$^d$\quad Computed as reported in section~\ref{other}.}\\
\end{tabular}
\label{tab:tableGX5-1_par}
\end{center}
\end{table}
%\footnote{in units of $10^{38}$}
%
%-----------------------------------
\subsection{GX~13$+$1}
\begin{figure*}

\hbox{\hspace{0.5cm}
\includegraphics[scale=0.35]{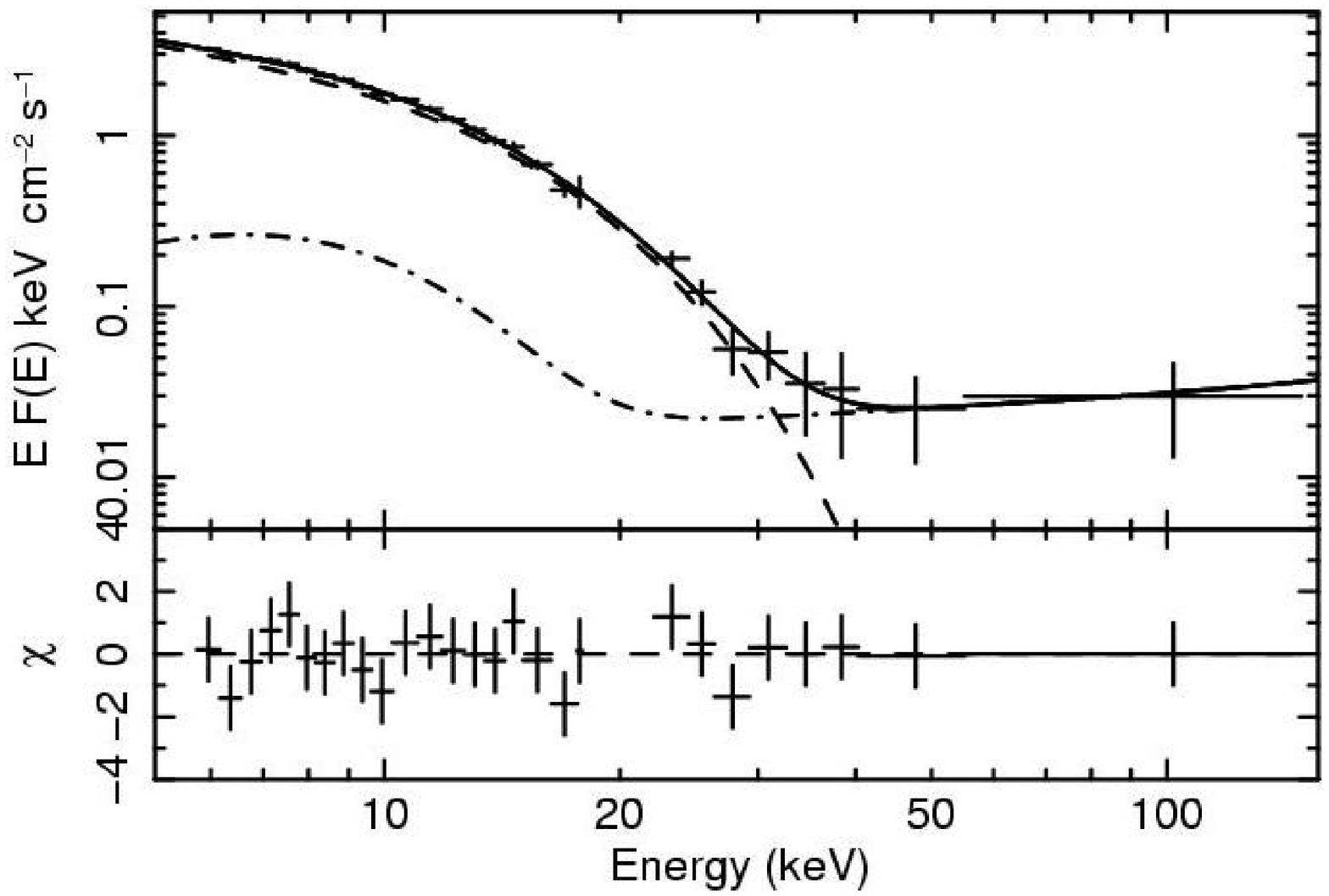}
\hspace{0.5cm}
\includegraphics[scale=0.35]{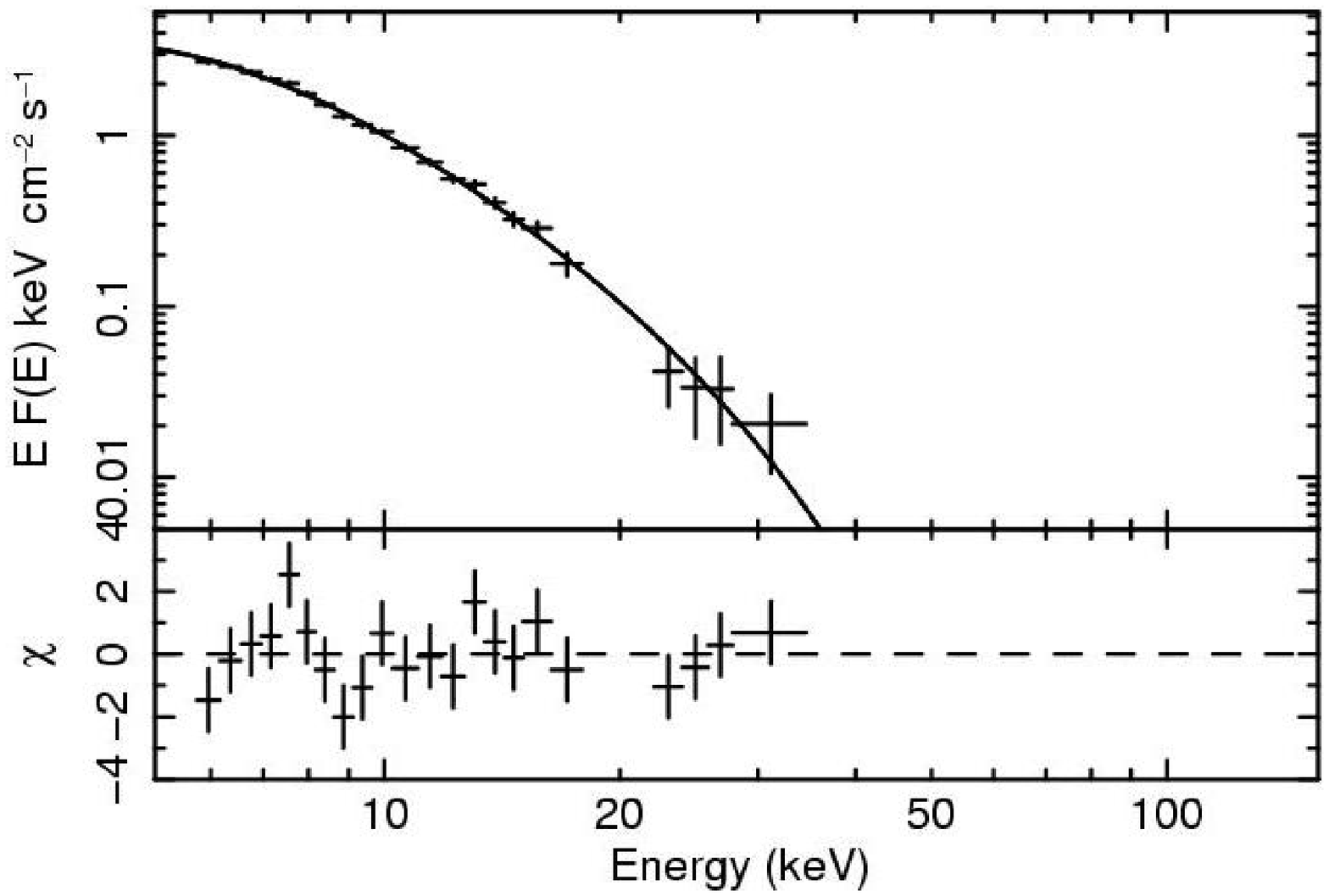}}
%\vbox{\vspace{0.15cm}}
%\hbox{\hspace{0.5cm}
%\includegraphics[scale=0.35]{./gx13p1_3_eeuf.ps}
%\hspace{0.5cm}}
%\vbox{\vspace{0.15cm}}
\caption[]{Unabsorbed EF(E) spectra of GX~13$+$1 for the HR intervals: HR$<$10 (left), 
HR$>$13 (right). The dotted lines are the two separated $\mathsf{compTB}$  components: dash for  $\mathsf{compTB[1]}$, dash-dot  $\mathsf{compTB[2]}$.} 
\label{fig:GX13+1eeuf}
\end{figure*}

Examining the shape of the  GX~13$+$1 JEM-X1 and IBIS/ISGRI spectra, we identified three intervals in which we recognized different phases of the spectral 
evolution and so we obtained three mean spectra %(Table~\ref{tab:tableGX13+1})
: the 
first mean spectrum corresponding to HR$<$10 shows a hard tail between 30 and 150\,keV characterizing an \textit{intermediate} state. A hard tail in the spectra of GX~13$+$1 had been discovered using \textit{INTEGRAL} data by P06. The mean spectrum 
corresponding to 10$<$HR$<$13 shows a \textit{high/soft} state, deprived of the hard tail, similar to the mean spectrum corresponding to HR$>$13.
Spectral analysis was carried out
%%has been accomplished 
using a double-component model only for the HR$<$10 spectrum, while the 
other two were fit with one \textit{thermal} $\mathsf{compTB}$ (see Table~\ref{tab:tableGX13+1_par}, Fig.~\ref{fig:GX13+1eeuf}). Looking at the parameters we note:
\begin{itemize}
\item the $\mathsf{compTB[1]}$ component, present in all spectra, describes the TC process. Again, for increasing values of HR we 
observe a strong decrease of CAF values and increase of $\alpha$, a signature of decreasing efficiency of the TC process;
\item the seed photon temperatures $kT_s$ in the two components are 
different in the first spectrum  (in the other two spectra only one \textit{thermal} $\mathsf{compTB}$ was needed); 
\item the plasma temperature $kT_e$ remains within the range of 3--4\,keV;
\item we applied the $\mathsf{compTB[2]}$ component only to  
the first spectrum as a \textit{mixed} $\mathsf{compTB}$: we detected a hard tail and an active BC process with $\delta\neq 0$; the $\alpha$ value was fixed to a value that corresponds 
to a minimum $\chi^2$ (see section \ref{results});
\item the luminosity of the blackbody $L_{bb}$ is very low with respect to the total \textit{X-ray}  luminosity of the source; the dimension of the blackbody emission region, 
$R_{bb}\sim$2\,km, is compatible with part of the NS surface and/or TL dimensions.
\end{itemize}
%
%\begin{table}
%  \begin{center}
%
%    \caption{The bright NS LMXB GX~13$+$1          }\vspace{1em}
%    \renewcommand{\arraystretch}{1.2}
%    \begin{tabular}[h]{cccc}
%      \hline
%Interval & pointings &\multicolumn{2}{c}{T$_{exp}$ (ksec)}   \\
%	& N  &  JEM-X  & ISGRI	\\	
%\hline
%\hline
%HR$<$10   &11 &29  &18      \\ 
%10$<$HR$<$13  &10 & 26 & 17    \\
%HR$>$13 &14  & 35 & 23   \\
%\hline
%
%      \end{tabular}
%    \label{tab:tableGX13+1}
%  \end{center}
%\end{table}
%
\begin{table}
\begin{center}
\caption{Best-fit parameters of the multi-component model $\mathsf{wabs\cdot(compTB[1]+compTB[2])}$ for the source  GX~13$+$1. Errors are computed at 90$\%$ confidence level for a single parameter.}
\begin{tabular}[h]{cccc}
\hline
\hline
parameter&HR$<$10&10$<$HR$<$13&HR$>$13\\
\hline
\multicolumn{4}{c}{$\mathsf{compTB[1]}$ (\,thermal: $\log A=8$, $\delta=0$\,)}\\
\hline
$kT_s$ $^a$& $0.2\,(<1.8)$&$0.92^{+0.07}_{0.08}$&$0.89^{+0.07}_{0.08}$\\
$kT_e$ $^a$&$3^{+0.4}_{-0.3}$&$3.5^{+1.2}_{-0.6}$&$4.2^{+7.5}_{-1.1}$\\
$\alpha$&$1.6^{+0.7}_{-0.5}$&$2.4^{+0.2}_{-0.4}$&$3^{+0.8}_{-0.5}$\\
$\tau$ $^d$&3.8&2.4&1.7\\
CAF $^d$&1.8&1.2&1.1\\
\hline
\multicolumn{4}{c}{$\mathsf{compTB[2]}$}\\
\hline
$kT_s$ $^a$&$1.7\,(>0.7)$&-&-\\
$R_{bb}$ $^{b,d}$&$2\,(<11)$&-&-\\
$kT_e$ $^a$&$[kT_e]$&-&-\\
$\alpha$&$[0.78]$&-&-\\
$\delta$&$103\,(>12)$&-&-\\
$\log A$&$-1.3\,(>-1.9)$&-&-\\
\hline
%$L_{tot}$ $^{c,d}$&1.2&0.6&0.5\\
%\hline
%$L_{bb}$ $^{c,d}$&0.03&-&-\\
$L_{tot}$ $^{c,d}$&1.8&0.6&0.5\\
\hline
$L_{bb}$ $^{c,d}$&0.03&-&-\\
&(1.7$\%$)&-&-\\
\hline
$\chi^2/dof$&14/20&32/19&22/19\\
\hline
\\
\multicolumn{4}{l}{$^a$\quad In keV.}\\
\multicolumn{4}{l}{$^b$\quad In km.}\\
\multicolumn{4}{l}{$^c$\quad In units of $10^{38}$\,erg\,sec$^{-1}$, in the 0.1--200\,keV energy range.}\\
\multicolumn{4}{l}{$^d$\quad Computed as reported in section~\ref{other}.}\\
\end{tabular}
\label{tab:tableGX13+1_par}
\end{center}
\end{table}
%--------------------------------------------
\subsection{GX~9$+$1}
\begin{figure}
\centering
\includegraphics[scale=0.35]{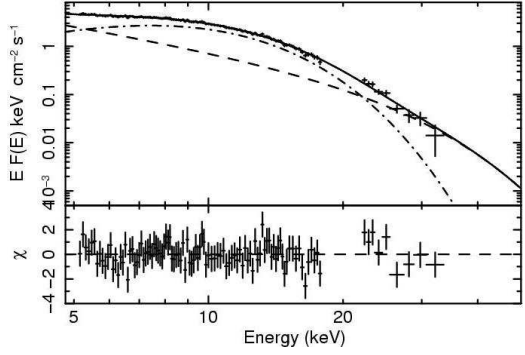}
\caption[]{Unabsorbed EF(E) spectrum for the HR interval 7.5$<$HR$<$8.5 for GX~9$+$1. The dotted lines are the two separated $\mathsf{compTB}$ components: dash for  $\mathsf{compTB[1]}$, dash-dot  $\mathsf{compTB[2]}$.} 
\label{fig:GX9+1eeuf}
\end{figure}

We classified every JEM-X1 and IBIS/ISGRI spectrum of GX~9$+$1 according to their shape and identified three HR intervals 
(HR$<$7.5, 7.5$<$HR$<$8.5 and HR$>$8.5) 
obtaining three
mean spectra.  %(Table~\ref{tab:tableGX9+1})
All spectra showed the \textit{high/soft} state, in fact, the spectral analysis 
was carried out 
with a double-component model using a \textit{thermal} $\mathsf{compTB}$ plus a second $\mathsf{compTB}$ as a pure \textit{blackbody} (see Table~\ref{tab:tableGX9+1_par}, Fig.~\ref{fig:GX9+1eeuf}):
\begin{itemize}
\item the TC process was 
described with $\mathsf{compTB[1]}$ in all three spectra; %they are very similar in shape but we can observe how the model cans distinguish and characterize the fine evolution differences: with increasing values of HR we have little decrement of the CAF and a little increasing of the $\alpha$ values;
\item similarly to the previous cases, the seed photon temperatures $kT_s$ 
differ in the two components:  
about 0.6\,keV for the first one and near 2\,keV for the second one;
\item in this case, the plasma temperature $kT_e$ is not well constrained by the data;
\item the presence of the second component $\mathsf{compTB[2]}$ reveals 
a constant blackbody contribution. Below 20\,keV the blackbody emission is more important than the TC one. 
The size of the blackbody-emitting region is nearly constant ($\sim$3.4 km), as is the corresponding X-ray luminosity ($0.2\times10^{38}$\,erg\,sec$^{-1}$). Despite the different HR selection, these spectra do not show a dramatic evolution, as shown in the previous cases.
 \end{itemize}  
%
% \begin{table}
%  \begin{center}
%
%    \caption{Bright NS LMXB GX~9$+$1          }\vspace{1em}
%    \renewcommand{\arraystretch}{1.2}
%    \begin{tabular}[h]{cccc}
%      \hline
%Interval & pointings  &\multicolumn{2}{c}{T$_{exp}$ (ksec)}   \\
%	& N& JEM-X & ISGRI	\\ 
%\hline
%\hline
%HR$<$7.5   & 19 & 50  & 34     \\ 
%7.5$<$HR$<$8.5  & 47 &119  & 79    \\
%HR$>$8.5 & 21  & 54  & 37   \\
%\hline
%
%      \end{tabular}
%    \label{tab:tableGX9+1}
%  \end{center}
%\end{table}
%
\begin{table}
\begin{center}
\caption{Best-fit parameters of the multi-component model $\mathsf{wabs\cdot(compTB[1]+compTB[2])}$ for the source GX~9$+$1. Errors are computed at 90$\%$ confidence level for a single parameter.}
\begin{tabular}[h]{cccc}
\hline
\hline
parameter&HR$<$7.5&7.5$<$HR$<$8.5&HR$>$8.5\\
\hline
\multicolumn{4}{c}{$\mathsf{compTB[1]}$ (\,thermal: $\log A=8$, $\delta=0$\,)}\\
\hline
$kT_s$ $^a$&$0.3\,(<0.6)$&$0.3\,(<0.6)$&$0.3\,(<0.6)$\\
$kT_e$ $^a$&$4^{+3}_{-1}$&$5^{+12}_{-1}$&$5\,(>3)$\\
$\alpha$&$2.1^{+0.6}_{-0.6}$&$2.4^{+0.7}_{-0.5}$&$2.4^{+0.9}_{-0.7}$\\
$\tau$ $^d$&2.5&2.0&1.9\\
CAF $^d$&1.5&1.4&1.4\\
\hline
\multicolumn{4}{c}{$\mathsf{compTB[2]}$ (\,blackbody: $\log A=-8$\,)}\\
\hline
$kT_s$ $^a$&$1.94^{+0.03}_{-0.04}$&$1.89^{+0.03}_{-0.03}$&$1.83^{+0.03}_{-0.05}$\\
$R_{bb}$ $^{b,d}$&$3.5^{+0.1}_{-0.1}$&$3.3^{+0.1}_{-0.1}$&$3.4^{+0.1}_{-0.1}$\\
$kT_e$ $^a$&-&-&-\\
$\alpha$&-&-&-\\
$\delta$&-&-&-\\
$\log A$&$[-8]$&$[-8]$&$[-8]$\\
\hline
%$L_{tot}$ $^{c,d}$&0.8&0.9&0.8\\
%\hline
%$L_{bb}$ $^{c,d}$&0.2&0.2&0.2\\
$L_{tot}$ $^{c,d}$&1.1&1.3&1.1\\
\hline
$L_{bb}$ $^{c,d}$&0.2&0.2&0.2\\
&(18$\%$)&(15$\%$)&(18$\%$)\\

\hline
$\chi^2/dof$&67/106&83/103&79/105\\
\hline
\\
\multicolumn{4}{l}{$^a$\quad In keV.}\\
\multicolumn{4}{l}{$^b$\quad In km.}\\
\multicolumn{4}{l}{$^c$\quad In units of $10^{38}$\,erg\,sec$^{-1}$, in the 0.1--200\,keV energy range.}\\
\multicolumn{4}{l}{$^d$\quad Computed as reported in section~\ref{other}.}\\
\end{tabular}
\label{tab:tableGX9+1_par}
\end{center}
\end{table}

%-----------------------------------------
 \subsection{GX~349$+$2}
\begin{figure}
\centering
%\hbox{\hspace{0.5cm}
\includegraphics[scale=0.35]{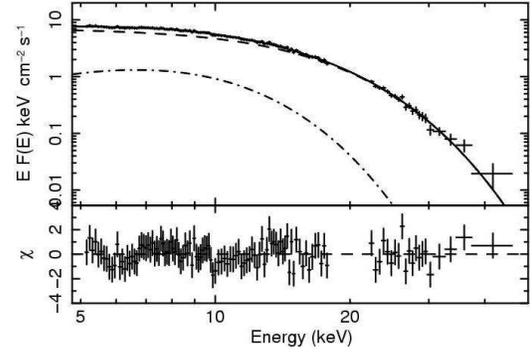}
%\hspace{0.5cm}
%\includegraphics[scale=0.35]{./gx349_2_eeuf.ps}}
%\vbox{\vspace{0.15cm}}
\caption[]{Unabsorbed EF(E) spectrum of GX~349$+$2 for  HR$<$7. 
The dotted lines are the two separated $\mathsf{compTB}$  components: dash for  $\mathsf{compTB[1]}$, dash-dot  $\mathsf{compTB[2]}$.} 
\label{fig:GX349+2eeuf}
\end{figure}

For GX~349$+$2, we identified two intervals (HR$<$7 and HR$>$7) showing similar spectra from which we obtained two mean spectra: both are of the \textit{high/soft} state well described 
with a double-component model consisting of a \textit{thermal} $\mathsf{compTB}$  and a second $\mathsf{compTB}$ as blackbody.
We  obtained the following results (see Table~\ref{tab:tableGX349+2_par}, Fig.~\ref{fig:GX349+2eeuf}):
\begin{itemize}
\item with $\mathsf{compTB[1]}$, we describe the TC process; for higher values of HR we do not observe the typical $\alpha$ growth but instead it is constant,  
which  brings to a constant CAF value and invariant TC efficiency;
\item similarly to previous sources, the seed photon temperatures are different 
when two $\mathsf{compTB}$ components are taken into consideration: $kT_s<1$\,keV for the \textit{thermal}  $\mathsf{compTB}$, higher temperature for the blackbody photons;
\item the plasma temperature stays always around 3\,keV;
\item using $\mathsf{compTB[2]}$, we detected the presence of a blackbody component: the blackbody luminosity $L_{bb}$ does not vary ($0.3\times10^{38}$\,erg\,sec$^{-1}$) and its emission region is of a dimension compatible with the NS and/or TL.
\end{itemize}
%
%\begin{table}
%  \begin{center}
%
%    \caption{Bright NS LMXB GX~349$+$2          }\vspace{1em}
%    \renewcommand{\arraystretch}{1.2}
%    \begin{tabular}[h]{cccc}
%      \hline
%Interval & pointings  &\multicolumn{2}{c}{T$_{exp}$ (ksec)}   \\
%	& N& JEM-X & ISGRI	\\	
%\hline
%\hline
%HR$<$7   & 20& 52 &  35    \\ 
%HR$>$7 & 31 & 82 &  53  \\
%\hline
%
%      \end{tabular}
%    \label{tab:tableGX349+2}
%  \end{center}
%\end{table}
%
\begin{table}
\begin{center}
\caption{Best-fit parameters of the multi-component model $\mathsf{wabs\cdot(compTB[1]+compTB[2])}$ for the source GX~349$+$2. Errors are computed at 90$\%$ confidence level for a single parameter.}
\begin{tabular}[h]{ccc}
\hline
\hline
parameter&HR$<$7&HR$>$7\\
\hline
\multicolumn{3}{c}{$\mathsf{compTB[1]}$ (\,thermal:$\log A=8$, $\delta=0$\,)}\\
\hline
$kT_s$&$0.3\,(<0.9)$&$0.4(<2.2)$\\
$kT_e$&$3^{+1.8}_{-0.1}$&$2.8^{+4.7}_{-0.08}$\\
$\alpha$&$1.1^{+0.2}_{-0.2}$&$1.1^{+0.5}_{-0.4}$\\
$\tau$ $^d$&5.0&5.2\\
CAF $^d$&2.1&2.0\\
\hline
\multicolumn{3}{c}{$\mathsf{compTB[2]}$ (\,blackbody: $\log A=-8$\,)}\\
\hline
$kT_s$&$1.7^{+0.3}_{-0.3}$&$1.2^{+0.7}_{-0.1}$\\
$R_{bb}$ $^{b,d}$&$5.2^{+2.6}_{-1.5}$&$9.8^{+2.5}_{-5.9}$\\
$kT_e$&-&-\\
$\alpha$&-&-\\
$\delta$&-&-\\
$\log A$&$[-8]$&$[-8]$\\
\hline
%$L_{tot}$ $^{c,d}$&3.0&2.0\\
%\hline
%$L_{bb}$ $^{c,d}$&0.3&0.3\\
$L_{tot}$ $^{c,d}$&3.4&2.2\\
\hline
$L_{bb}$ $^{c,d}$&0.3&0.3\\
&(9$\%$)&(14$\%$)\\
\hline
$\chi^2/dof$&104/115&76/114\\
\hline
\\
\multicolumn{3}{l}{$^a$\quad In keV.}\\
\multicolumn{3}{l}{$^b$\quad In km.}\\
\multicolumn{3}{l}{$^c$\quad In units of $10^{38}$\,erg\,sec$^{-1}$, in the 0.1--200\,keV energy range.}\\
\multicolumn{3}{l}{$^d$\quad Computed as reported in section~\ref{other}.}\\
\end{tabular}
\label{tab:tableGX349+2_par}
\end{center}
\end{table}
%-----------------------------------------
\subsection{GX~3$+$1}
\begin{figure}
\centering
\includegraphics[scale=0.35]{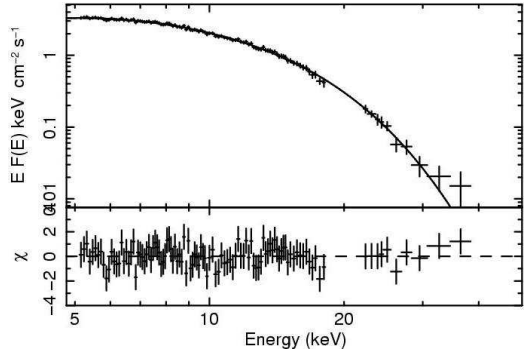}
\caption[]{Unabsorbed EF(E) spectrum for the HR interval 
HR$<$8 for GX~3$+$1. The overall spectrum is well described by a single component, $\mathsf{compTB[1]}$, see text.} 
\label{fig:GX3+1eeuf}
\end{figure}

In the case of GX~3$+$1, we identified three intervals of HR in which we recognized different phases of spectral evolution
in the JEM-X1 and 
IBIS/ISGRI spectra: the three mean spectra obtained (HR$<$8, 8$<$HR$<$9.5 and HR$>$9.5) all represent
the \textit{high/soft} state.
The spectral fitting was 
carried out on all spectra with a single component model, a \textit{thermal}  $\mathsf{compTB}$
%% , for all the spectra
(see Table~\ref{tab:tableGX3+1_par}, Fig.~\ref{fig:GX3+1eeuf}):   
\begin{itemize}
\item as usual with $\mathsf{compTB[1]}$ we describe the TC process; similarly to previous sources, for increasing HR values, we see a decreasing of the TC efficiency (decrease of CAF and increase of $\alpha$);
\item the seed photon temperature $kT_s$ is of the order 1\,keV;
\item the plasma temperature $kT_e$ is of the order 3--4\,keV.
\end{itemize}
%
%\begin{table}
%  \begin{center}
%
%    \caption{The bright NS LMXB GX~3$+$1          }\vspace{1em}
%    \renewcommand{\arraystretch}{1.2}
%    \begin{tabular}[h]{cccc}
 %     \hline
%Interval & pointings  &\multicolumn{2}{c}{T$_{exp}$ (ksec)}   \\
%	& N& JEM-X & ISGRI	\\	      
%\hline
%\hline
%HR$<$8   &37 & 107 & 70     \\ 
%8$<$HR$<$9.5  &103 & 323 & 210    \\
%HR$>$9.5 & 26 & 78 & 52   \\
%\hline
%
%      \end{tabular}
%    \label{tab:tableGX3+1}
%  \end{center}
%\end{table}
%
\begin{table}
\begin{center}
\caption{Best-fit parameters of the model $\mathsf{wabs\cdot(compTB[1])}$ for the source GX~3$+$1. Errors are computed at 90$\%$ confidence level for a single parameter.}
\begin{tabular}[h]{cccc}
\hline
\hline
parameter&HR$<$8&8$<$HR$<$9.5&HR$>$9.5\\
\hline
\multicolumn{4}{c}{$\mathsf{compTB[1]}$ (\,thermal: $\log A=8$, $\delta=0$\,)}\\
\hline
$kT_s$ $^a$&$0.9^{+0.09}_{-0.13}$&$1.04^{+0.06}_{-0.06}$&$1.13^{+0.07}_{-0.06}$\\
$kT_e$ $^a$&$2.7^{+0.1}_{-0.1}$&$3^{+0.1}_{-0.2}$&$4^{+1.3}_{-0.7}$\\
$\alpha$&$1.4^{+0.2}_{-0.2}$&$1.9^{+0.2}_{-0.2}$&$2.8^{+0.5}_{-0.4}$\\
$\tau$ $^c$&4.4&3.2&1.9\\
CAF $^c$&1.4&1.2&1.1\\
\hline
%$L_{tot}$ $^{b,c}$&0.3&0.3&0.3\\
$L_{tot}$ $^{b,c}$&0.3&0.2&0.2\\
\hline
$\chi^2/dof$&65/106&116/114&100/108\\
\hline
\\
\multicolumn{4}{l}{$^a$\quad In keV.}\\
\multicolumn{4}{l}{$^b$\quad In units of $10^{38}$\,erg\,sec$^{-1}$, in the 0.1--200\,keV energy range.}\\
\multicolumn{4}{l}{$^c$\quad Computed as reported in section~\ref{other}.}\\
\end{tabular}
\label{tab:tableGX3+1_par}
\end{center}
\end{table}
%
%\begin{figure*}

%\hbox{\hspace{0.5cm}
%%\includegraphics[scale=0.35]{./gx3p1_1_eeuf.ps}
%\hspace{0.5cm}
%\includegraphics[scale=0.35]{./gx3p1_2_eeuf.ps}}
%%\includegraphics[scale=0.35]{./gx3p1_3_eeuf.ps}}
%\vbox{\vspace{0.15cm}}
%\hbox{\hspace{0.5cm}
%\includegraphics[scale=0.35]{./gx3p1_3_eeuf.ps}
%\hspace{0.5cm}}
%\vbox{\vspace{0.15cm}}
%\caption[]{Unabsorbed EF(E)spectra of GX~3$+$1 for the HR intervals: HR$<$8 (left), 
%HR$>$9.5 (right).} 
%\label{fig:GX3+1eeuf}
%\end{figure*}

\section{Discussion} 
In this work we have studied the spectral evolution of five NS LMXBs using the new Comptonization model $\mathsf{compTB}$ \citep{farinelli08}.
The $\mathsf{compTB}$  model is formed by two components: one component is a pure blackbody directly seen by the observer while the other one takes into account the contribution of thermal and bulk Comptonization of the blackbody seed photons.
To fit the \textit{INTEGRAL} broad band spectra we applied two $\mathsf{compTB}$ models.

Every source has shown a different spectral evolution: for GX~5--1 and GX~13$+$1 increasing values of the HR correspond to spectral changes from the \textit{intermediate} state (high-energy hard tail), to the \textit{high/soft} state (no hard tail) while in the other three sources we did not observe dramatic spectral changes despite a progressive softening  for increasing accretion rate (i.e. HR, see below).

In all sources, regardless of HR,  the total energetic budget is dominated by the thermal Comptonization component of $kT_s \lesssim 1$\,keV BB-like seed photons off warm electrons  with temperature $kT_e$ typically of 3--4\,keV and optical depth of a few ($\tau \lesssim 5$). This component,  described with a \textit{thermal} $\mathsf{compTB}$ (equivalent to a $\mathsf{compTT}$) is likely associated to the region
between the geometrically thin accretion disk and the NS surface, the so-called Transition Layer, where most of the gravitational energy release takes place \citep[][F08]{titarchuk04}.\\
When a second $\mathsf{compTB}$ model is required by the data, a second hotter seed photon population ($\sim$\,1--2\,keV) can be identified, interpreted as the BB-like contribution, from the innermost region of the system, the NS and the inner TL (Fig.~\ref{fig:TL}).

% \begin{figure}
%\centering
%%\includegraphics[width=1.0\linewidth]{/nereide/laura/Desktop/laura/immagini/TL_eng.ps}
%\includegraphics[width=1.0\linewidth]{./COMPTB_articolo.ps}
%%\vspace{4cm}./image_ticks.ps
%\caption{Schematic view of the proposed scenario for thermal and bulk Comptonization regions in LMXBs hosting NSs.
%\label{fig:TL}}
%\end{figure} 

A rough estimation of the apparent emitting radius of this BB region provides in fact results which are consistent with a NS and/or TL radius, with  $R_{bb}\sim$\,2--12\,km. 
In the case of \mbox{GX~5--1} for which we have the best statistics, we can see (Table~\ref{tab:tableGX5-1_par}) an increase of the emitting BB region with the HR (hence $\dot M$, see next section). 

Where a single $\mathsf{compTB}$ model has been applied, we observe that $kT_s$ has values very near to 1\,keV, unlike the other cases where $kT_s$ is either below this energy ($\lesssim$\,0.6\,keV) or higher (always $>1$\,keV).
In the single $\mathsf{compTB}$ model case, it is likely that a part of the seed photons come from the disk (very soft photons $\lesssim$0.5\,keV) but emission from the TL and NS surface can contaminate the seed photon population increasing its overall temperature ($\sim$1\,keV).
As none of the observed sources reached the \textit{hard state} (see, e.g. Fig.~2 in P06) where an increased height-scale of the TL may provide a high geometrical coverage of the central compact object, the lack of a second component in the case of a single $\mathsf{compTB}$ component in the fit is likely due to poor statistics in addition to small energy coverage in the band where most of the BB-like emission is present.

In terms of the observable quantities which may trace the source spectral evolution, it is important to focus on the $\alpha$-parameter, related to the slope of the Green's function of the Comptonization region. As already pointed out in F08, for \textit{intermediate state} sources, we are dealing in fact with two observable indexes, one related to the pure TC process, while the second one related to the TC+BC efficiency. The first index, only depends on the system temperature and optical depth and is actually directly related to the Comptonization parameter $y$\footnote{The Comptonization parameter is  defined as \mbox{$y = \frac{kT_e}{m_ec^2}\max(\tau_e,\tau_e^2)$} where the \textit{optical depth} of the scattering region in NS LMXBs is typically $\tau_e^2\gg 1$.}, while the second index contains an additional dependence on the bulk parameter $\delta$ 
(see Eq.[23] in TMK97). In both cases, the slope of the Green's function depends on the first eigenvalue of the spatial problem for
a bounded medium. In fact, when BC is present, it is possible somewhat to treat the two regions
(internal bulk-dominated TL, external TC dominated TL plus innermost accretion disk) in a separate way, each one with its own spatial configuration and associated eigenvalue problem and derived spectral index. This translates into two different observable quantities (the thermal index $\alpha$, plus the TC+BC index).  For systems where BC is suppressed, or strongly reduced, $\delta$ goes to zero, and according to Eq.[13] in TMK97, the two indexes tend to be coincident and one single
TC component is observed.

The $\alpha$ parameter measures the slope of the spectrum and can quantify the effective efficiency of the overall Comptonization processes i.e. how much energy has been exchanged between the photon field and the electrons. In the first $\mathsf{compTB}$ (the thermal one), $\alpha$ becomes higher as HR increases: we can always observe the spectral evolution in the direction of a progressive softening, a smaller quantity of energy is gained by the photon field reducing the maximum energy of the spectra. 
Indeed, the energy gained by thermal Comptonization, CAF, decreases as $\alpha$ increases, giving a physical evaluation of the energy gain through the TC process.
The $\alpha$ value related to the \textit{mixed} $\mathsf{compTB}$ is generally higher with respect to the \textit{thermal} one. It encloses both the BC and TC contributions coming from the TL, the NS itself and possibly the inner accretion disk.
\subsection{The transient nature of the hard tail}\label{transient}
Bulk Comptonization acts in the TL, where the orbits of the in-falling matter deviate from the Keplerian behaviour to adjust to the NS rotation speed. 
Here the matter, during its bulk motion ($v_b\sim 0.1c$, Titarchuk $\&$ Farinelli in prep.), interacts with the photon field that 
originates in the inner disk, in the TL itself and from the impact of matter onto the NS.\\
Since the physical conditions of the interaction region strongly influence the BC, this produces the \textit{transient} high energy hard tail behaviour. 
Indeed the local accretion rate of the source determines the resulting radiation pressure in the inner TL 
and in turn influences its hydrodynamical configuration. 

At very high values of accretion rate, the radiation pressure behaves like an intense photon wind \citep{bradshaw07} able to break efficiently the accreting matter, abating the $v_b$, hence  
suppressing (or strongly reducing) the Fermi first-order process in photon energy gain which is responsible for
generation of the hard X-ray tail.
%%to the photon field ($\Delta E_b \propto v_b$, \textbf{Reference?}) and the consequence is a cut-off at lower energies ($\delta$ decreases) . 
In the $\mathsf{compTB}$ model, the quantity that allows us to evaluate the hard tail-$\dot M$ link is the bulk parameter $\delta$ which is defined as
\begin{equation}
\delta \equiv \frac{\left\langle \Delta E_{Bulk} \right\rangle} {\left\langle \Delta E_{th} \right\rangle}= \frac{1-l}{\dot m \Theta}\label{delta}
\end{equation}
where $l \equiv L/L_{Edd}$  and  $\dot m \equiv \dot M/\dot M_{Edd}$ are  \textit{local} luminosity and accretion rate in Eddington's units, respectively,  and 
$\Theta \equiv kT_e/m_ec^2$ is the adimensional electron temperature\footnote{We note that the correct expression of the $\delta$ parameter is the one 
written in the text. In \cite{farinelli08}, Eq.[2] erroneously indicates a square root in the right term.}.\\
Increasing the accretion rate leads the system to a state  closer to thermal equilibrium between electrons and the background radiation field,  through a more  efficient Compton cooling of the electrons themselves (lower average energy gain of the photons and CAF decreasing).

In this work this spectral evolution has been observed for increasing values of HR with a slow spectral softening starting from the \textit{intermediate} 
state (low HR) to the \textit{high/soft} state (high HR values). Given the spectral evolution observed in all the sources, we can reasonably associate the 
increase of HR to an increasing accretion rate. 
As already pointed out by previous authors \citep[e.g.][]{homan07, lin09, vanderklis01}, the role of $\dot M$ in the spectral evolution of these sources is
not straightforward. Furthermore, from the observational results of a detailed time-resolved \textit{BeppoSAX} spectral analysis of the Z source \mbox{Cyg~X--2}, F09 proposed to split the general definition of the accretion rate $\dot M$, into two quantities,  $\dot M_{\rm disk}$ and $\dot M_{\rm TL}$, obeying the condition $\dot M_{\rm TL} \la  \dot M_{\rm disk}$. The reason for this resides in the fact that
the total measured source bolometric luminosity does not follow the expected $\dot M$ trend from the source position in the hardness intensity diagram, and thus cannot be used as a tracer of the \textit{total} mass accretion rate. Moreover, as also reported in \cite{bradshaw07}, for high (Eddington-like) accretion rates, a strong radiation pressure may originate in the accretion disk, which may
eventually eject part of the accreting material, so that the mass flow at the disk/TL radius is
actually lower than the one flowing trough the Roche lobe. In this picture, the key role in the innermost system is played by $\dot M_{\rm TL}$.

We can thus read the observed spectral variability of our sources as the interplay of TC and BC processes according to the aforementioned considerations and increasing HR:  
\begin{enumerate}
\item for a low $\dot M_{TL}$, the hard tail results to be absent or too weak because of a lack of a sufficiently high number of BC scatterings able to elevate the hard tail contribution over the sensitivity instrument threshold. The observed spectrum is due mainly 
to the TC of  soft seed photons from the disk and NS in the hot corona plasma (tens of keV). This is the case of low-dim Atolls not considered here (see F08);
\item for higher $\dot M_{TL}$, the corona starts to be efficiently cooled and the overall spectrum is dominated by TC in a cold corona (kT$_{e}$ of a few keV).
BC may not be strong enough to be detected (GX~3$+$1, GX~9$+$1, GX~349$+$2, but see also point 4); 
\item for increasing $\dot M_{TL}$, we can observe the \textit{intermediate} state (GX~5--1, GX~13$+$1 in this work), in which the hard 
tail is visible. Its importance and intensity are strongly influenced by the accretion rate value. Indeed only within a narrow  range of $\dot M_{TL}$, 
the number of BC interactions is high enough to produce the hard tail and at the same time the radiation pressure is not too high to reduce the BC efficiency. 
The fine tuning of these two quantities could be the reason why we observe a little number of sources hosting NS displaying hard tails, and even more, only 
a given spectral state within a given source (the \textit{intermediate} state, corresponding to the so-called Horizontal Branch of Z sources);
\item in case of very high values of $\dot M_{TL} $, the hard tail disappears or its contribution results to be negligible being under the sensitivity threshold. 
The pressure in the TL increases because a large amount of matter releases a lot of energy at the impact onto the NS or within the TL 
itself and this can inhibit the bulk motion. 
The source moves to the \textit{high/soft} spectral state in which the spectrum is mainly defined by the TC process plus 
eventually a variable contribution of a blackbody component emitted from the TL and/or the NS surface (GX~3$+$1, GX~9$+$1, GX~349$+$2).  
\end{enumerate}
In the case of GX~5--1 for which we have the best statistics and HR evolution, we can relate the increasing trend of the $R_{bb}$ value (size of the 
BB emitting region) to the increasing values of HR/$\dot M $: at large values of local accretion rates, i.e. radiation pressure, 
bulk motion is inhibited and the matter falls at lower speed, following a complex behaviour for which it may arrive on 
the NS surface spreading over a larger area \citep{inogamov99}. The large amount matter covering the NS can produce a layer able to suppress the radio emission (P06).
\subsection{The absence of the hard tail in most bright Atoll "GX" sources}
From the spectral states observed and the information obtained by the fitting process, it is difficult to say whether the permanent absence of hard X-ray tail in sources like \mbox{GX~3$+$1}, \mbox{GX~9$+$1} is due to a \textit{locally} high-enough accretion rate which is able to suppress BC or in fact to the opposite case, where BC is present but too low to be detected. In these sources such a transient feature has in fact never been observed, even with the past missions. 

In GX 5-1, a transient hard X-ray tail was instead previously detected by \textit{Ginga} \citep{asai94} and \textit{INTEGRAL} \citep{paizis05}, in GX 13+1 it was detected for the first time by \textit{INTEGRAL} (P06), while in GX~349$+$2, the feature was observed by \textit{BeppoSAX} \citep{disalvo01} when the source was out-of-flare. Also taking into account previous results on other Z sources (GX~17$+$2, Sco~X--1, GX~340$+$0, see e.g. P06 for a complete list of references) it is currently difficult to provide a global self-consistent picture which may explain this
variate phenomenology. We still do not have an observable parameter which may help to unambiguously determine, at least phenomenologically, which is the threshold for triggering the hard X-ray tail.

In the accretion scenario where the production of high-energy photons is inhibited by high \textit{local} levels of the accretion rate (bulk stopping through radiation pressure), one should
look with particular care at the contribution of the BB-like emission to the total luminosity. 
If this BB-like emission is due to energy release of the accreting matter both to the NS surface and
in the TL (through viscous dissipation), one would expect that hard tail quenching would be correlated to higher percentage of the BB contribution to the total luminosity. 

Using $L_{\rm bb}/L_{\rm tot}$ as
a possible tracer of the \textit{local accretion rate}, and thus of the BC efficiency, would have the advantage of being distance-independent, a not secondary issue when comparing the estimated luminosity among sources.
Such a quantity may however be significantly different among sources which show a hard X-ray tail, ranging from $\sim$2\% (in GX~13$+$1, this work) to $\sim$25\% in Cyg~X--2 (F08). Of course, a satisfactory energy coverage below 1\,keV is of prime importance to give unbiased estimations of this contribution.
In this sense the reported values of $L_{\rm tot}$ and $L_{\rm bb}/L_{\rm tot}$ in the present work must be treated very cautiously, as the energy band covered by JEM-X1 starts from 5\,keV, thus above the threshold where most of BB energy is emitted.

With this prescriptions in mind, following the spectral evolution of GX~5--1 with increasing HR, we note that the presence of the hard X-ray tail is anti-correlated with the total source luminosity
and the percentage of the energetic budget carried-out by the BB-like component, even though the
latter change is marginal (see Fig.~\ref{fig:GX5-1eeuf} and  Table~\ref{tab:tableGX5-1_par}).
The other sources for which we can detect two components in the spectrum along the observed HR range, but with no bulk contribution detected, unlike for GX~5--1, are GX~9$+$1  and GX~349$+$2 (see Figs.~\ref{fig:GX9+1eeuf} and \ref{fig:GX349+2eeuf}). In this case case the BB-like component contributes about 20\% and 10\% of the source luminosity, respectively.
Little can be said though for GX~3$+$1 or GX~13$+$1 for which we do not have the double component evolution along with HR.
\section{Conclusions}
Despite an increasing amount of theoretical and observational results, the picture giving rise to the presence of variable PL-like hard X-ray emission in NS LMXBs is far from being completely clear.
Investigating the physical conditions of the innermost region of these sources is very critical. The bulk Comptonization approach, together 
with the physical scenario proposed,  can be considered an important step forward in trying to understand what is at the origin of the transient hard 
tail behaviour. Nevertheless, it is currently not easy to satisfactorily explain why  sources like   GX~9$+$1 and GX~5--1 that do have a similar 
X-ray spectrum and related parameters (kT$_{e}$, kT$_{s}$ of the disk and NS-TL, $\tau$) yet differ for such an important feature as the 
presence of the transient hard tail. 

Even if we were able to quantify $\dot M_{TL}$ (disentangling it from the overall $\dot M$) we would face the problem of 
understanding what makes $\dot M_{TL}$ change in the different cases i.e. what is the origin of the inflow anisotropy. In fact, a theoretical and/or observational quantity which may unambiguously trigger or dump the hard X-ray emission has not been identified yet. Of course it is possible that we are dealing with a multi-parametric problem.

A possible highly speculative explanation could 
reside in the different composition of the disk: in Atoll sources the orbital period is in general relatively short (\lsim 5 hr)
and in the Z sources for which it is known it is long (\gsim 10 hr). This would indicate that Z sources
may contain an evolved companion star \citep[e.g. \mbox{Cyg~X--2}, GX~1$+$4][and references therein]{liu07} while Atolls contain a main sequence star \citep[e.g. Aql~X--1, Ser~X--1][and references therein]{liu07}. This  is consistent with the fact that the number of known Atoll sources 
(25) is higher than the number of known Z sources \citep[8,][]{liu07} since
the main sequence phase of a star is much longer than the one of an evolved star making
the detection of Atolls more likely \citep{verbunt95}. The accretion disk coming from an evolved companion (Z-like system) would be Helium-rich (i.e. heavier) hence a higher 
luminosity would be required to stop the bulk flow, making the detection of BC more likely.

The chemical composition of the disc could be inferred by the duration of 
type I X-ray bursts: long bursts are thought to be due to mixed H/He 
burning, triggered by thermally unstable He ignition, expected at inferred 
near-Eddington accretion rates \citep{kuulkers07}. 
However the observed behaviour is far from being interpreted within the 
current bursting theory due to the presence of unexpected short burst or even 
absence thereof, in a regime where long bursts would be expected \citep{kuulkers07}. 
The reason of this could reside in the fact that in the case of high  
accretion rates, accretion is unlikely through a plain disc 
and the TL configuration may also have an effect on the bursting 
behaviour.

Additional or concurring effects could reside in the presence of a counter-rotating NS with respect to the Keplerian disk rotation, whose main effect could be the presence of a higher
level of gravito-kinetic energy release in the TL, in the region where the angular velocity
adjusts from the Keplerian regime to that of the slowly spinning NS. Such higher energy release
would in turn produce a higher local radiation pressure gradient, more efficient in stopping bulk.
Unfortunately, this is highly speculative and currently difficult to prove. A first approach to tackle the problem would be 
a more detailed study of the spectral evolution together with the long-term behaviour of these sources (Savolainen et al, in prep.).

Understanding the origin of the hard tails in NS LMXBs is important to make a decisive
step forward, from phenomenology (power-laws) to physics. This step forward will not make
us understand hard tails alone, on the contrary, hard tails could be the observational
feature that gives us the means to understand the mechanism of accretion flows in general, see e.g \cite{shrader98}  and \cite{ferrigno09} for an application of bulk motion Comptonization in Black Hole LMXBs and High Mass X-Ray Binaries, respectively.
%
%===================================================
\begin{acknowledgements}
Based on observations with \textit{INTEGRAL}, an ESA project
with instruments and science data centre funded by ESA member states
(especially the PI countries: Denmark, France, Germany, Italy,
Spain, and Switzerland), Czech Republic and Poland, and with the
participation of Russia and the USA.\\
LM, AP and RF acknowledge L. Titarchuk for useful discussions.
AP acknowledges the Italian Space Agency financial and 
programmatic support via contract I/008/07/0. 
This work has been partially supported by the grant from Italian 
PRIN-INAF 2007, "Bulk motion Comptonization models in X-ray Binaries: 
from phenomenology to physics", PI M. Cocchi.

\end{acknowledgements}

\bibliographystyle{aa}
\bibliography{biblio}

\end{document}